\documentclass{article}

\usepackage{spconf,amsmath,graphicx}
\usepackage{url}  % Tuan add for url in ref
\usepackage{balance} % Tuan add for balancing columns in latex IEEE pub
\usepackage{url}  % Tuan add for url in ref

\usepackage[linesnumbered,lined,ruled,norelsize]{algorithm2e}
\usepackage{setspace}

\DeclareMathOperator*{\argmin}{argmin}  % Tuan add for argmin

\makeatletter
\def\@oddfoot{\footnotesize Copyright 2017 IEEE. Published in the IEEE 2017 International Conference on Image Processing (ICIP 2017), scheduled for 17-20 September 2017 in Beijing, China. }
\makeatother

\title{360-degree Video Stitching for Dual-fisheye Lens Cameras Based On Rigid Moving Least Squares}
\name{Tuan Ho$^{\star}$ \qquad Ioannis D. Schizas$^{\star}$ \qquad K. R. Rao$^{\star}$ \qquad Madhukar Budagavi$^{\dagger}$}

\address{$^{\star}$ Dept. of Electrical Engineering, University of Texas--Arlington, Arlington, TX USA \\
	  $^{\dagger}$ Samsung Research America, Richardson, TX USA}

\begin{document}
%IEEEoverridecommandlockouts 
%IEEEpubid{makebox[columnwidth]{978-1-4673-9486-4/16/$31.00~copyright~2016 IEEE hfill} hspace{columnsep}makebox[columnwidth]{ }}

%\setlength{\textfloatsep}{15pt} % change the space between top float and text
%\setlength{\intextsep}{15pt} % change the space between top float and text
\setlength{\textfloatsep}{10pt} % change the space between top float and text
\setlength{\intextsep}{10pt} % change the space between top float and text
% \textfloatsep: distance between top float to the text below
% \intextsep: distance between 'HERE' float to the above and below text

%%%%%%%%%%%%%%%%%%%%%%%%%%%%%%%%%%%%%%%%%%%%%%%%%%%%%%%%%%%%%%%%
%%%%%%%%%%%%%%  Path to the Images/Figures    %%%%%%%%%%%%%%%%%%
%%%%%%%%%%%%%%%%%%%%%%%%%%%%%%%%%%%%%%%%%%%%%%%%%%%%%%%%%%%%%%%%
\graphicspath{{../Figures/}}
	
%\ninept
%
%\IEEEoverridecommandlockouts
%\footnotetext{Copyright 2017 IEEE. Published in the IEEE 2017 International Conference on Image Processing (ICIP 2017), scheduled for 17-20 September 2017 in Beijing, China. Personal use of this material is permitted. However, permission to reprint/republish this material for advertising or promotional purposes or for creating new collective works for resale or redistribution to servers or lists, or to reuse any copyrighted component of this work in other works, must be obtained from the IEEE. Contact: Manager, Copyrights and Permissions / IEEE Service Center / 445 Hoes Lane / P.O. Box 1331 / Piscataway, NJ 08855-1331, USA. Telephone: + Intl. 908-562-3966.}

%\IEEEpubid{\Copyright\ 2017 IEEE. Published in the IEEE 2017 International Conference on Image Processing (ICIP 2017), scheduled for 17-20 September 2017 in Beijing, China.}

\maketitle

\begin{abstract}
Dual-fisheye lens cameras are becoming popular for 360-degree video capture, especially for User-generated content (UGC), since they are affordable and portable. Images generated by the dual-fisheye cameras have limited overlap and hence require non-conventional stitching techniques to produce high-quality 360x180-degree panoramas. This paper introduces a novel method to align these images using interpolation grids based on rigid moving least squares. Furthermore, jitter is the critical issue arising when one applies the image-based stitching algorithms to video. It stems from the unconstrained movement of stitching boundary from one frame to another. Therefore, we also propose a new algorithm to maintain the temporal coherence of stitching boundary to provide jitter-free 360-degree videos. Results show that the method proposed in this paper can produce higher quality stitched images and videos than prior work.
\end{abstract}
%

%%%%%%%%%%%%%%%%%%%%%%%%%%%%%%%%%%%%%%%%%%%%%%%%%%%%%%%%%%%%%%%%%%%
%%%%%%                  INTRODUCTION                      %%%%%%%%%
%%%%%%%%%%%%%%%%%%%%%%%%%%%%%%%%%%%%%%%%%%%%%%%%%%%%%%%%%%%%%%%%%%%

\section{Introduction}
\label{sec:intro}

%%%%%%%%%%%%%%
Dual-fisheye lens cameras are becoming popular for 360-degree video capture, especially for UGC. Their portability and affordability give them an edge over traditional and professional 360-degree capturing systems such as \cite{gopro_odyssey}\cite{facebook_surround360} which usually deploy 6--17+ cameras on the same rig and are very expensive to own. An example dual-fisheye lens 360-degree camera is the Samsung Gear 360 which can produce 360x180-degree panoramas and video that are viewable on the 360-degree viewers such as Cardboard \cite{google_cardboard} or GearVR \cite{samsung_gearvr}. Other examples of dual-fisheye lens 360-degree cameras are Ricoh Theta \cite{Ricoh_Theta} and LG 360 Cam \cite{LG_360} to name a few.
%% -------------------------------------------------------------------------
%\begin{figure}[!b]
%	\begin{minipage}[b]{1.0\linewidth}
%		\centering
%		\centerline{\includegraphics[width=\textwidth]{360_0533_Cam_1_small_size.jpg}}
%	\end{minipage}
%	%
%	\caption{The images (7776 x 3888 pixels) taken by Gear 360 dual-fisheye lens camera with buildings and patterned backgrounds on the stitching boundaries.}
%	\label{fig:Outdoor_Pattern}
%	%
%\end{figure}
%% -------------------------------------------------------------------------

However, the convenience of a compact and affordable 360-degree capture system comes with a caveat. The images generated by the dual-fisheye lenses have limited overlap, and as we show in \cite{tuan:ICASSP2017}, the conventional stitching methods such as ones in \cite{brown_IJCV07} \cite{Szeliski:CV} do not provide satisfactory stitching results.

%\footnotetext{ Copyright 2017 IEEE. Published in the IEEE 2017 International Conference on Image Processing (ICIP 2017), scheduled for 17-20 September 2017 in Beijing, China. }

\begin{figure}[t]
	\begin{minipage}[b]{1.0\linewidth}
		\centering
		\centerline{\includegraphics[width=\linewidth]{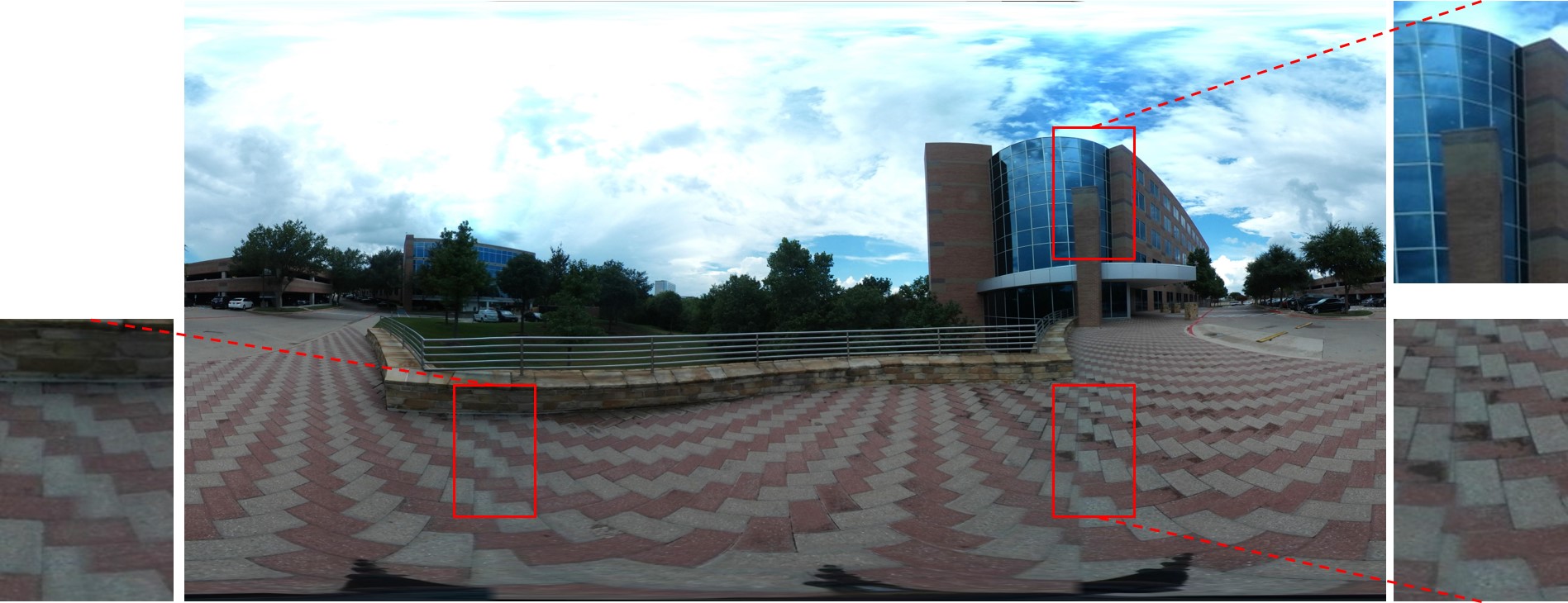}}
	\end{minipage}
	%
	%	\caption{360x180-degree panorama stitched by \cite{tuan:ICASSP2017} (from the fisheye images in \figurename~\ref{fig:Outdoor_Pattern}) and the discontinuities in the overlapping regions.}
	\caption{360x180-degree panorama stitched by \cite{tuan:ICASSP2017} and the discontinuities in the overlapping regions.}
	\label{fig:ICASSP_stitch}
\end{figure}

\begin{figure*}[t]
	\begin{minipage}[b]{1.0\linewidth}
		\centering
%		\centerline{\includegraphics[width=\linewidth]{ICIP_ICASSP_flow_black_noFill.png}}
		\centerline{\includegraphics[width=\linewidth]{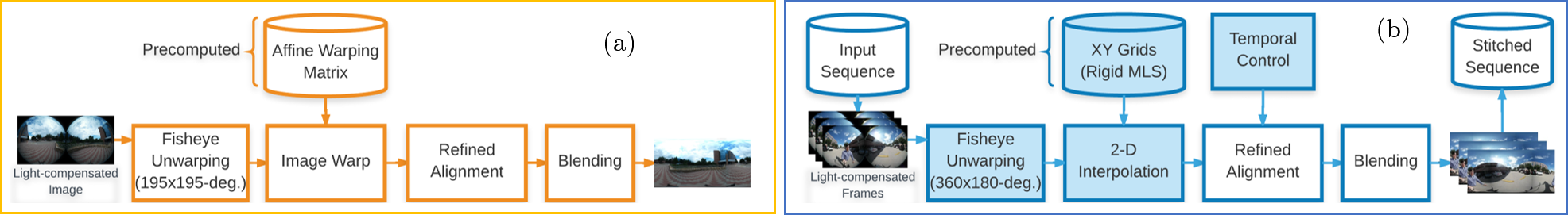}}
	\end{minipage}
	\caption{The processing flow \cite{tuan:ICASSP2017} (a) and the approach of this paper (b).}
	\label{fig:Flow_ICIP_ICASSP}
\end{figure*}

To stitch the images generated by the dual-fisheye lens cameras, \cite{tuan:ICASSP2017} suggests a framework of four main stages, as shown in \figurename~\ref{fig:Flow_ICIP_ICASSP}(a). The first and second stages compensate for the light-fall off of the fisheye-lens camera and transform the light-compensated fisheye images into an equirectangular format that can be viewed on 360-degree players respectively. After the first two stages, the fisheye-unwarped images does not align with each other. Therefore, \cite{tuan:ICASSP2017} proposes a two-step registration method that minimizes the discontinuity in the overlapping regions to align the images and blend them together. In this approach, the first step compensates for the geometric misalignment between the two fisheye lenses and depends on the camera parameters. The second step is a more refined alignment that adjusts any discontinuities caused by objects with varying depth in the stitching boundaries. In the first alignment, \cite{tuan:ICASSP2017} solves an over-determined system for a warping matrix which is then used to align the images. This results in a least-squares approximated solution which globally transforms the images. Our observation is that typically the control points in the central part of the 360-degree image get aligned well resulting in improved quality compared to prior techniques. However, the control points at the top and bottom part of the image do not get aligned precisely leading to stitching artifacts in those regions. \figurename~\ref{fig:ICASSP_stitch} shows an example of visible discontinuities in the stitching boundary of pictures with patterns on the background.

% -------------------------------------------------------------------------
%
% -------------------------------------------------------------------------

This paper builds upon our previous work in \cite{tuan:ICASSP2017} and improves the image alignment and stitching performance over the entire stitched 360x180-degree panoramas. It uses rigid moving least squares approach to achieve the improved alignment. This paper also extends the work to video stitching by incorporating a new temporal-coherent algorithm to produce jitter-free 360-degree videos.

% -------------------------------------------------------------------------
% Image spreads 2 columns
%
% -------------------------------------------------------------------------

%%%%%%%%%%%%%%%%%%%%%%%%%%%%%%%%%%%%%%%%%%%%%%%%%%%%%%%%%%%%%%%%%%%
%%%%%%             THE PROPOSED ALGORITHM                 %%%%%%%%%
%%%%%%%%%%%%%%%%%%%%%%%%%%%%%%%%%%%%%%%%%%%%%%%%%%%%%%%%%%%%%%%%%%%

\section{The proposed algorithm}
\label{sec:algorithm}

%Similar to \cite{tuan:ICASSP2017}, the proposed alignment has two steps. The first step compensates for the geometric misalignment between the two fisheye lenses in the camera based on our novel approach using rigid moving least squares (MLS). The second step is a more refined alignment that adjusts any small discontinuities caused by varied object's depth in the stitching boundaries. \figurename~\ref{fig:Flow_ICIP_ICASSP} show the processing flow of \cite{tuan:ICASSP2017} and this proposal.

\figurename~\ref{fig:Flow_ICIP_ICASSP}(b) shows the block diagram of the proposed algorithm in this paper. Similar to \cite{tuan:ICASSP2017}, the proposed image alignment also has two steps -- the first one is dependent on camera parameters, and the second step works adaptively to the scene. However, instead of estimating a warping matrix in a least-squares sense to align the pictures in the first step, we generate interpolation grids to deform the image based on rigid moving least squares (MLS) approach.

\subsection{Rigid Moving Least Squares}
\label{ssec:MLS}

Let $p$ and $q$ be the control points in the overlapping regions of the original and deformed images respectively. \cite{Schaefer:SIGGRAPH06} defines three properties of an image deformation function $f$ which are: interpolation ($f(p_i) = q_i$ under deformation), smoothness (preserves smooth transition among pixels), and identity ($q_i = p_i \Rightarrow f(v) = v$). 

For every point in the image, we solve for a transformation matrix $M$ that minimizes the weighted least squares:

\vspace{-0.02in}
%-----------------------------------Equation----------------------%
\begin{equation}
\label{eq:eqn_1}
\argmin\limits_{M}\sum\limits_{i} w_i\,\Big\|{\hat{p_i}M - \hat{q_i}}\Big\|^2  %% 
\end{equation}

%%% \argmin\limits_{M}\sum\limits_{i} w_i\norm{\hat{p_i}M - \hat{q_i}}^2  %% norm is not supported in \package{amsmath} but {mathtools} ---------   \|{}\| is equivalent to \norm{} in mathtools

%------------------------------------------------------------------%
\vspace{-0.02in}

\noindent where  the weights $w_i$ are proportional to the distance between the image point $v$ and the control point $p_i$ in the sense that $w_i$ gets smaller when $v$ moves further away from $p_i$ (i.e. the least squares minimization depends on the point of evaluation, thus the name moving least squares). When $v \to p_i$, $f$ interpolates $f(v) = q_i$. \cite{Schaefer:SIGGRAPH06} defines such weights as:

\vspace{-0.05in}

$$ w_i = \frac{1}{\displaystyle |p_i - v|^{2\alpha}} $$

\noindent $\hat{p_i} = v - p^{*}$ and $\hat{q_i} = q^{*}$ are derived from each point $v$ in the image and the weighted centroids $p^{*}$ and $q^{*}$ \cite{Schaefer:SIGGRAPH06}.

%$$ p^{*} = \frac{\sum_i{w_i p_i}}{\sum_i{w_i}} \space, \qquad q^{*} = \frac{\sum_i{w_i q_i}}{\sum_i{w_i}} $$

%\vspace{0.05in}

\begin{figure}[t]
	\begin{minipage}[b]{1.0\linewidth}
		\centering
		\centerline{\includegraphics[width=\textwidth]{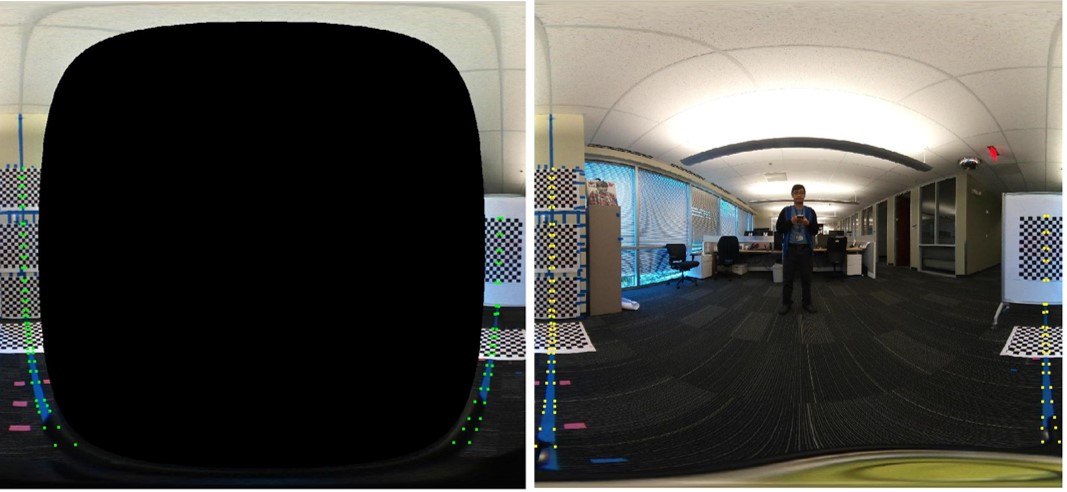}}
	\end{minipage}
	\caption{Left: the overlapping areas on the unwarped left image and $\{q\}_i$ (green dots). Right: unwarped right image and $\{p\}_i$ (yellow dots).}
	\label{fig:Control_Points}
\end{figure}

For control points selection, we adopted the checkerboard experiment from \cite{tuan:ICASSP2017} with our method of picking the correspondence points. In this experiment, both fisheye lenses, each has 195-degree field of view, see the same checkerboards on their sides. The images taken by the fisheye lenses are unwarped to 360x180-degree equirectangular planes. We then arrange the unwarped images so that the right image is positioned at the center of the 360x180-degree plane, while the left image is split and put to the sides of the plane. With this arrangement, the overlapping regions are ready for control-point selection. By choosing the same checkerboards' cross sections on the unwarped images, one can visualize the geometric misalignment between the two lenses. \figurename~\ref{fig:Control_Points} shows the selected control points $\{p\}_i$ and $\{q\}_i$, which indicate the differentiated positions of the same points in the stitching boundaries of the two images. Our interest is to determine the function $f_r$ that does the transformation $f_r(p_i)=q_i$ in the overlapping regions while keeping the other areas of the image as visually intact as possible.

While the MLS is general in the matrix $M$ in \eqref{eq:eqn_1}, we are only interested in the rigid transformation since it generates more realistic results than affine and similarity transformations. The similarity transformations are a subset of the affine transformations that have only translation, rotation, and uniform scaling. The similarity matrix M is defined such that $M^TM = \lambda^2I$ (e.g. a rotation matrix). $\lambda^2$ acts as a uniform scaling factor. In rigid transformation, it is desirable that no uniform scaling is included. \cite{Schaefer:SIGGRAPH06} proposed a theorem that relates the MLS solution for $M^TM = \lambda^2I$ (similarity transformation) to its solution of $M^TM = I$ (rigid transformation), and derived the solution for the rigid MLS function $f_r$. We invite readers to read \cite{Schaefer:SIGGRAPH06} for more details about the mathematical treatment used here.

%\vspace{-0.20in}
%
%%-----------------------------------Equation----------------------%
%\begin{equation}
%\label{eq:eqn_2}
%M = \frac{1}{\displaystyle\mu_r}\sum\limits_{i} w_i \begin{pmatrix}
%\hat{p_i} \\
%-\hat{p_i}^\perp{} \\
%\end{pmatrix}  %% 
%\begin {pmatrix}
%\hat{q_i}^T & -\hat{q_i}^{\perp T} \\
%\end{pmatrix}
%\end{equation}
%%------------------------------------------------------------------%
%
%\noindent where $\mu_r$ is a constant in \cite{Schaefer:SIGGRAPH06} and $\perp$ is a 2-D operator that rotates the target point by 90 degrees (i.e. $(x,y)^\perp = (-y,x)$). The rigid MLS function $f_r(v)$ is described as:
%
%%\vspace{-0.15in}
%
%%%-----------------------------------Equation----------------------%
%%\begin{equation}
%%\label{eq:eqn_3}
%%\mu_r = \sqrt{\Bigg(\sum\limits_{i} w_i \hat{q_i} \hat{p_i}^T\Bigg)^2 + \Bigg(\sum\limits_{i} w_i \hat{q_i} \hat{p_i}^{\perp T}\Bigg)^2}
%%\end{equation}
%%%------------------------------------------------------------------%
%
%%\vspace{0.02in}
%
%%\noindent and $\perp$ is a 2-D operator that rotates the target point by 90 degrees (i.e. $(x,y)^\perp = (-y,x)$). The rigid MLS function $f_r(v)$ is described as:
%
%\vspace{-0.06in}
%
%%-----------------------------------Equation----------------------%
%\begin{equation}
%\label{eq:eqn_4}
%f_r(v) = | v - p^{*} | \frac{\vec{f_r}(v)}{|\vec{f_r}(v)|} + q^{*}
%\end{equation}
%%------------------------------------------------------------------%
%%\vspace{-0.02in}
%%\noindent where \quad $\vec{f_r}(v) = (v - p^{*}) \, M $
%%\vspace{-0.07in}
%%
%%$$ \vec{f_r}(v) = (v - p^{*}) \, M $$
%
%%\vspace{0.02in}

\begin{figure}[t]
	\begin{minipage}[b]{1.0\linewidth}
		\centering
		\centerline{\includegraphics[width=\textwidth]{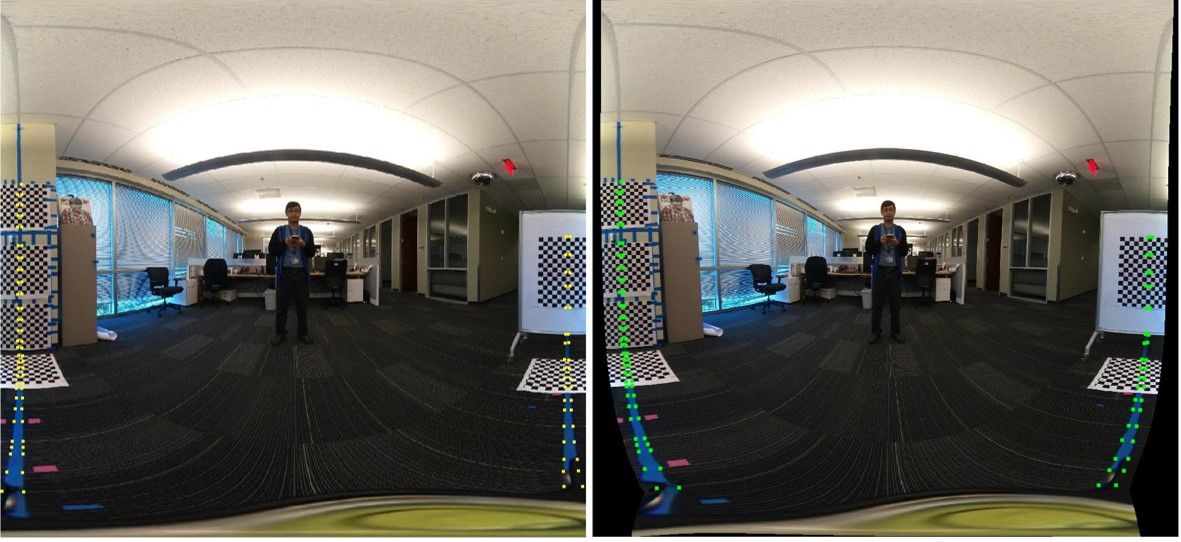}}
	\end{minipage}
	\caption{The original and the rigid-MLS-deformed images with their control points $\{p\}_i$ and $\{q\}_i$ overlayed.}
	\label{fig:MLS_Rigid_Deform}
\end{figure}

We generate the rigid-MLS interpolation grids to deform the right unwarped image (i.e. to apply $f_r$ over the image), thus aligning it with the left one. \figurename~\ref{fig:MLS_Rigid_Deform} shows the right unwarped fisheye image gets deformed by the rigid MLS method. While the portions of the image in proximity to the stitching boundaries are transformed to match the other image, the remaining of the deformed picture have no discernible difference compared to the original.

%%%%%%%%%%%%%%%%%%%%%%%%%%%%%%%%%%%%%%%%%%%%%%%%%%%%%%%%%%%%%%%%%%%

\subsection{Refined Alignment}
\label{ssec:Refine_Align}

The rigid MLS aligns the control points around the stitching boundary, thus registering the two unwarped fisheye images together. However, when the depth of the object in the overlapping areas changes, it introduces misalignment to the scene. Therefore, a refined alignment is necessary after the rigid MLS deformation. To this end, we adopt the same adaptive method of using the normalized cross-correlation matching in \cite{tuan:ICASSP2017} to further align the images.

\begin{figure}[t]
	\begin{minipage}[b]{1.0\linewidth}
		\centering
		\centerline{\includegraphics[width=\textwidth]{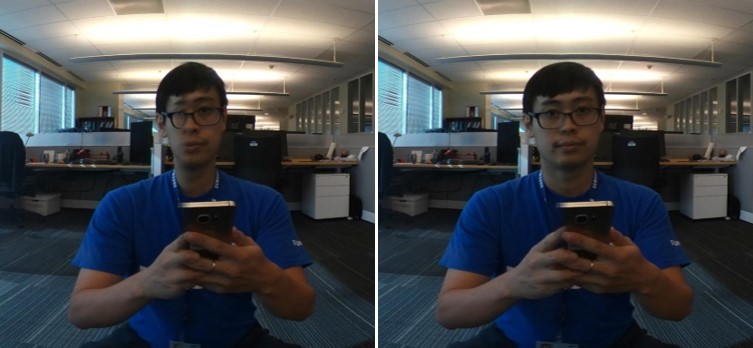}}
	\end{minipage}
	\caption{The person sitting close to the camera and in the stitching boundary. Left: after rigid MLS deformation. Right: after rigid MLS deformation and refined alignment.}
	\label{fig:XCorr}
\end{figure}

The refined alignment performs a fast template matching and utilizes the matching displacements on both stitching boundaries to generate eight pairs of control points. These points are then used to solve for a 3x3 affine matrix to warp the deformed image. As a least-squares solution, this refined method is not sufficient in registering images with complicated misalignment patterns, but it works very well for those with minor misalignment such as the one caused by the varied object's depth. \figurename~\ref{fig:XCorr} shows that the refined alignment minimizes the discontinuity when the person is sitting close to the camera's stitching boundary.

%%%%%%%%%%%%%%%%%%%%%%%%%%%%%%%%%%%%%%%%%%%%%%%%%%%%%%%%%%%%%%%%%%%

%\subsection{Unwarping Method}
%\label{ssec:Unwarping}
%
%Fisheye unwarping is the process of transforming the fisheye distorted images into a natural appearance format when viewed on 360-degree viewers. \cite{tuan:ICASSP2017} discussed a fisheye unwarping that maps a $195$x$195$-degree field of view (fov) fisheye image into a $195$x$195$-degree equirectangular one. The proposed method, however, maps $195$x$195$-degree fisheye image to $360$x$180$-degree equirectangular to avoid suppression (i.e. straight lines not straight after unwarping) near the edge of the fisheye image, which has the most optical distortion. This is an approach that is very similar to the one in \cite{Panotools}. \figurename~\ref{fig:ICIP_Stitch}(c) shows that the method used in this proposal can produce a little less-distorted unwarped images, thus making it easier for the alignment process in generating well-stitched panoramas.

% -------------------------------------------------------------------------
% -------------------------------------------------------------------------

% -------------------------------------------------------------------------
% -------------------------------------------------------------------------

% -------------------------------------------------------------------------
% -------------------------------------------------------------------------

% -------------------------------------------------------------------------
% Figure Spread two columns
%
\begin{figure*}[t]
	\begin{minipage}[b]{1.0\linewidth}
		\centering
		\centerline{\includegraphics[width=\linewidth]{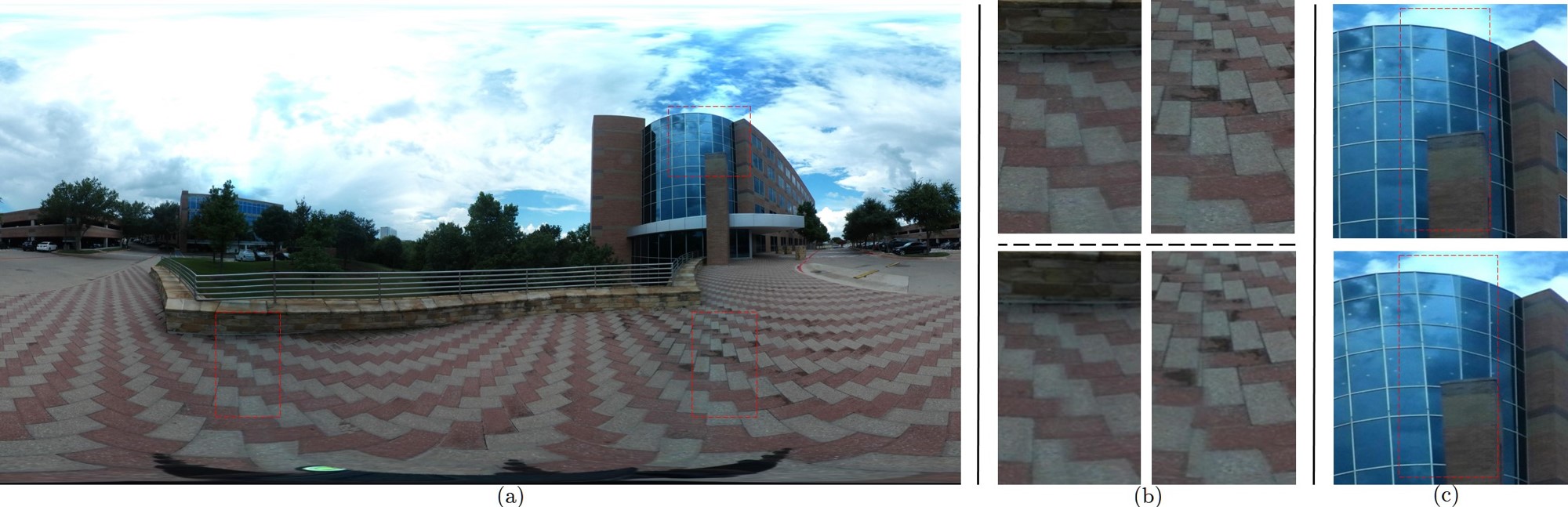}}
	\end{minipage}
	\caption{(a) 360x180-degree panorama stitched by this proposal. (b)(c) The stitching boundaries in (a) (top row) compared to the same image stitched by \cite{tuan:ICASSP2017} (bottom row).}
	\label{fig:ICIP_Stitch}
\end{figure*}

% -------------------------------------------------------------------------

% -------------------------------------------------------------------------
% Figure Spread two columns
%
\begin{figure*}[!h]
	\begin{minipage}[b]{1.0\linewidth}
		\centering
		\centerline{\includegraphics[width=\linewidth]{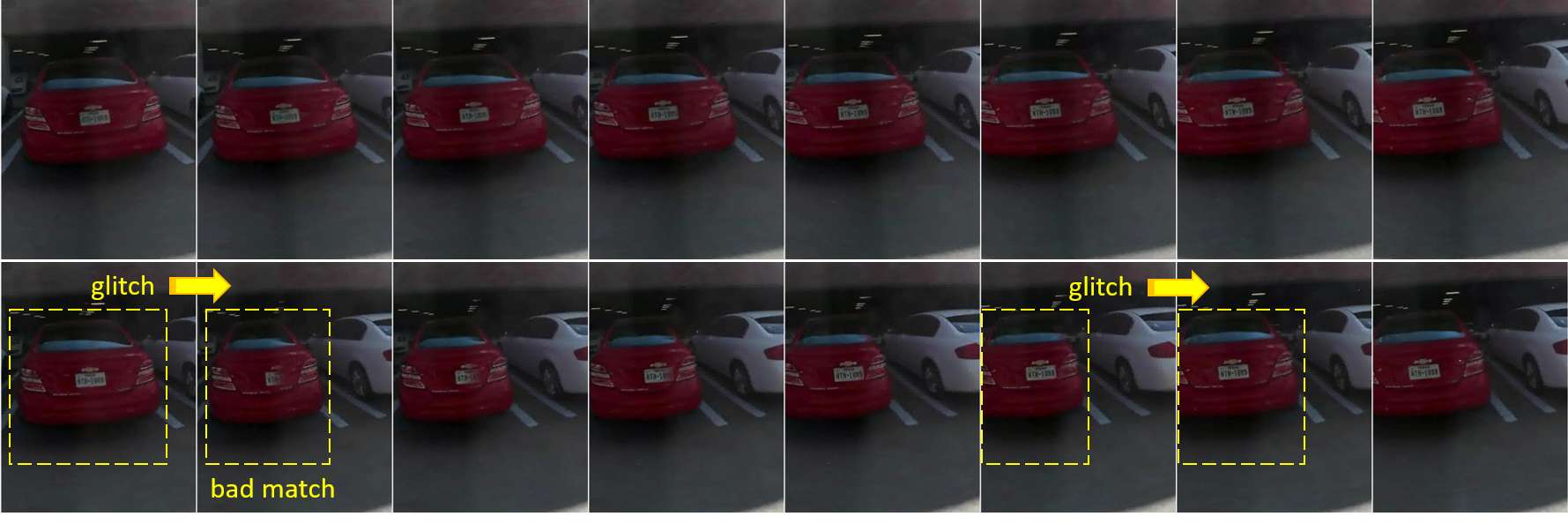}}
	\end{minipage}
	\caption{Stitching boundary in consecutive frames. Stitched by the proposal (top row), and by \cite{tuan:ICASSP2017} (bottom row).}
	\label{fig:video_ICIP_vs_ICASSP}
\end{figure*}
% -------------------------------------------------------------------------

%%%%%%%%%%%%%%%%%%%%%%%%%%%%%%%%%%%%%%%%%%%%%%%%%%%%%%%%%%%%%%%%%%%
%%%%%%             360-VIDEO                              %%%%%%%%%
%%%%%%%%%%%%%%%%%%%%%%%%%%%%%%%%%%%%%%%%%%%%%%%%%%%%%%%%%%%%%%%%%%%

\section{Extension to 360-degree video}
\label{sec:360video}

In the 360-degree video stitching, it is essential to minimize jitters--the abrupt transition between the stitched frames so that the final video appears continuous and comfortable to view. Adjacent frames in the sequence that are not stitched by the same measure can generate jitters. In the work presented here, when a bad match occurs without getting filtered out in the refined alignment, it generates a false warping matrix that abruptly distorts the stitching boundary of the picture. This attenuated scene causes jitter which is the result of the sudden transition between the previous well-stitched frame and the current bad-stitched one. Therefore, it is important to guarantee good matches throughout the entire sequence to maintain smooth frame-to-frame transition, and thus minimize jitters.

\begin{algorithm}[!h]
%	\setstretch{1.0}
%	\DontPrintSemicolon
	
	\SetKwData{DtscoreLeft}{scoreLeft}
	\SetKwData{DtscoreRight}{scoreRight}
	\SetKwData{DtwarpEn}{warpEn} \SetKwData{DtaffineMat}{affineMat}
	\SetKwData{DtRefineAvai}{scoreLeft}
	
	\SetKwFunction{FnTemplateMatch}{TemplMatch}
	\SetKwFunction{FnAffMatEst}{AffineMatEstimate}
	\SetKwFunction{FnWarpImg}{WarpImg}
	
	\KwIn{$leftImage$, $rightImage$ (deformed) }

	(\DtscoreLeft, \DtscoreRight) $\leftarrow$ \FnTemplateMatch{}\;

	\eIf{$($ both matching scores are good $)$} 
	  {
	  	Estimate affine warping matrix \DtaffineMat\;
	  	Store \DtaffineMat for the next frame\;
	  	\DtwarpEn $\leftarrow 1$\;
	  } (\tcp*[h]{bad scores on either boundary})
	  {
	  	\eIf{ matching scores of the previous frame are good }
	  	{
		  	\DtwarpEn $\leftarrow 1$\;
		  	\DtaffineMat $\leftarrow previous$ \DtaffineMat\; 
	  	}
	  	{
		  	\DtwarpEn $\leftarrow 0$\tcp*[l]{don't warp image}	
	  	}

	  } % matching good

	\If{$($ \DtwarpEn $)$}
	{
		Warp $rightImage$ by \DtaffineMat\;
	}
	\caption{Refined Alignment (with jitter control)}\label{algo:refineAlign}
\end{algorithm}

Algorithm~\ref{algo:refineAlign} illustrates our method to maintain the temporal coherence for the sequence. A good score is returned at one stitching boundary if all of the followings satisfied. First, the peak normalized cross-correlation is larger than $0.85/1.0$. Second, the returned vertical displacement is in the margin of $[-10,+10]$ pixels. Third, the horizontal displacement of the current match must not exceed $10\%$ margin compared to its of the previous frame. These constraints, obtained from our empirical experiments, are set to eliminate bad matching caused by poor lighting and abrupt movements of the boundaries in horizontal and vertical directions.

%%%%%%%%%%%%%%%%%%%%%%%%%%%%%%%%%%%%%%%%%%%%%%%%%%%%%%%%%%%%%%%%%%%
%%%%%%             IMPLE & RESULTS                        %%%%%%%%%
%%%%%%%%%%%%%%%%%%%%%%%%%%%%%%%%%%%%%%%%%%%%%%%%%%%%%%%%%%%%%%%%%%%

\section{Implementation and Results}
\label{sec:results}

We have implemented the proposal algorithm in C++ and Matlab. The rigid MLS grids are precomputed, and the deformation becomes an interpolation process that can be accelerated by GPU.

\figurename~\ref{fig:ICIP_Stitch}(a) illustrates an image stitched by the proposed method. In this picture, there are a fence, buildings, and patterned background on the stitching boundaries. \figurename~\ref{fig:ICIP_Stitch}(b) shows the comparison of the stitching boundaries in the image stitched by the proposal and by \cite{tuan:ICASSP2017} (also in \figurename~\ref{fig:ICASSP_stitch}). While the discontinuities appear in the stitching boundaries in \cite{tuan:ICASSP2017} as the result of the least-squares solution, the proposed method produces seamless 360-degree panorama thanks to the rigid MLS deformation.

%Not only the proposed method can produce better stitched images compared to the prior work, it is also able to produce jitter-minimized 360-degree videos.
For video stitching\footnotemark, \figurename~\ref{fig:video_ICIP_vs_ICASSP} demonstrates the adjacent stitched frames created by the proposal (top row, no jitter) and by \cite{tuan:ICASSP2017} (bottom row). In the bottom row, the first jitter occurs when the refined alignment lets a bad match get through, resulting in an affine transformation that moves the image on the right side of the stitching boundary to the left. As a result, the car in the boundary gets distorted leading to an abrupt transition between the frames.

\footnotetext{This paper has supplementary downloadable materials, which are the stitched videos generated by this novel method and by \cite{tuan:ICASSP2017}.}

%%%%%%%%%%%%%%%%%%%%%%%%%%%%%%%%%%%%%%%%%%%%%%%%%%%%%%%%%%%%%%%%%%%
%%%%%%             CONCLUSION                             %%%%%%%%%
%%%%%%%%%%%%%%%%%%%%%%%%%%%%%%%%%%%%%%%%%%%%%%%%%%%%%%%%%%%%%%%%%%%

\section{Conclusion}

This paper has introduced a novel method for stitching the images and video sequences generated by the dual-fisheye lens cameras. The proposed alignment has two steps. The first one, carried out offline, compensates for the sophisticated geometric misalignment between the two fisheye lenses on the camera based on rigid moving least squares approach. The second step, applied online and adaptively to the scene, provides a more refined adjustment for any misalignment created by the objects with varying depth  on the stitching boundaries. We extend the proposed approach to 360-degree video stitching with the relevant constraints to maintain the smooth transition between frames and therefore minimize jitters. Results show that our method not only generates more accurately stitched 360x180-degree images but also jitter-free 360-degree videos.

% Below is an example of how to insert images. Delete the ``\vspace'' line,
% uncomment the preceding line ``\centerline...'' and replace ``imageX.ps''
% with a suitable PostScript file name.
% -------------------------------------------------------------------------
%\begin{figure}[htb]
%
%\begin{minipage}[b]{1.0\linewidth}
%  \centering
%  \centerline{\includegraphics[width=8.5cm]{image1}}
%%  \vspace{2.0cm}
%  \centerline{(a) Result 1}\medskip
%\end{minipage}
%%
%\begin{minipage}[b]{.48\linewidth}
%  \centering
%  \centerline{\includegraphics[width=4.0cm]{image3}}
%%  \vspace{1.5cm}
%  \centerline{(b) Results 3}\medskip
%\end{minipage}
%\hfill
%\begin{minipage}[b]{0.48\linewidth}
%  \centering
%  \centerline{\includegraphics[width=4.0cm]{image4}}
%%  \vspace{1.5cm}
%  \centerline{(c) Result 4}\medskip
%\end{minipage}
%%
%\caption{Example of placing a figure with experimental results.}
%\label{fig:res}
%%
%\end{figure}

%\vfill
%\pagebreak

%%%%%%%%%%%%%%%%%%%%%%%%%%%%%%%%%%%%%%%%%%%%%%%%%%%%%%%%%%%%%%%%%%%
%%%%%%                  REFERENCES                        %%%%%%%%%
%%%%%%%%%%%%%%%%%%%%%%%%%%%%%%%%%%%%%%%%%%%%%%%%%%%%%%%%%%%%%%%%%%%
\clearpage
%\newpage

\bibliographystyle{IEEEbib}
\bibliography{strings,refs}

\balance % Tuan add for balancing columns in ref.

\end{document}